\documentclass[twocolumn,showpacs,preprintnumbers,amsmath,amssymb]{revtex4}

\usepackage{graphicx}
\usepackage{dcolumn}
\usepackage{bm}

\newcommand{\bra}[1]{\langle #1 |}
\newcommand{\ket}[1]{| #1 \rangle}
\newcommand{\braket}[2]{\langle #1\,| #2\rangle}

\newcommand{\norm}[1]{| #1 |}
\newcommand{\acom}[2]{\left\{#1,#2\right\}}
\newcommand{\com}[2]{\left[#1,#2\right]}
\newcommand{\ca}[1]{a^\dag_{#1}}
\newcommand{\da}[1]{a_{#1}}

\newcommand{\cb}[1]{b^\dag_{#1}}
\newcommand{\db}[1]{b_{#1}}
\newcommand{\cB}[1]{B^\dag_{#1}}
\newcommand{\dB}[1]{B_{#1}}
\newcommand{\cF}[1]{F^\dag_{#1}}
\newcommand{\dF}[1]{F_{#1}}
\newcommand{\cbB}[1]{\bar{B}^\dag_{#1}}
\newcommand{\dbB}[1]{\bar{B}_{#1}}
\newcommand{\calf}[1]{\alpha^\dag_{#1}}
\newcommand{\dalf}[1]{\alpha_{#1}}
\newcommand{\cP}[1]{C^\dag_{#1}}
\newcommand{\dP}[1]{C_{#1}}
\newcommand{\cbP}[1]{\bar{C}^\dag_{#1}}
\newcommand{\dbP}[1]{\bar{C}_{#1}}

\newcommand{\la}[4]{\lambda\!\left(_{#1\ #3}^{#2\ #4}\right)}
\newcommand{\kvec}[1]{{\bf #1}}
\newcommand{\clos}[2]{I^{(#1)}_{#2}}

\newcommand{\gamn}[1]{\gamma^{(#1)}_N}
\newcommand{\tgamn}[1]{\tilde{\gamma}^{(#1)}_N}
\newcommand{\ggamn}[1]{\Gamma^{(#1)}_N}

\newcommand{\comb}[2]{{\cal C}_{#1}^{#2}}
\newcommand{\pix}[2]{\pi^{(X)}_{#1\,#2}}
\newcommand{\pin}[2]{\pi^{(\nu)}_{#1\,#2}}
\newcommand{\pim}[2]{\pi_{#1\,#2}}
\newcommand{\snk}[2]{S^{(#2)}_{#1}}
\newcommand{\hypergF}[3]{F(#1,#2;#3)}
\newcommand{\hypergG}[3]{G(#1,#2;#3)}

\begin{document}

\preprint{APS/123-QED}

\title{Closure relations for composite bosons: difference between polaritons and Wannier or Frenkel excitons}

\author{M. Combescot}
 \email{Monique.Combescot@insp.jussieu.fr}
\affiliation{%
Institut des NanoSciences de Paris, Universit\'e Pierre et Marie Curie, CNRS, 140 rue de Lourmel, 75015 Paris, France}%

\author{M.A. Dupertuis}
\affiliation{
Laboratoire d'Opto\'electronique Quantique et Laboratoire de Physique des Nanostructures, Ecole Polytechnique F\'ed\'erale de Lausanne EPFL,
Station 3, CH-1015 Lausanne, Switzerland
 \email{Marc-Andre.Dupertuis@epfl.ch}
}%

\date{\today}

\begin{abstract}
We derive the closure relation for $N$ polaritons made of three different types of excitons: bosonized excitons, Frenkel or Wannier excitons. In the case of polaritons made of Wannier excitons, we show how this closure relation, which appears as non-diagonal, may reduce to the one of $N$ elementary bosons, the photons, with its $1/N!$ prefactor, or to the one of $N$ Wannier excitons, with its $(1/N!)^2$ prefactor. Widely different forms of closure relations are thus found depending on the composite bosons at hand. Comparison with closure relations of excitons, either bosonized or kept composite as Frenkel or Wannier excitons, allows us to discuss the influence of a reduction of the number of internal degrees of freedom, as well as the importance of the composite nature of the particles and the existence of fermionic components. 
\end{abstract}


\maketitle

\section{Introduction}

Although it is widely claimed that an even number of fermions behaves as a boson, this cannot be fully correct due to a very fundamental reason: replacing a pair of free fermions by an elementary boson is a drastic alteration of phase space. This must in particular reflect strongly through closure relations. We have actually shown~\cite{CBM_PRB05,CBMDub_PhysRep07} that the closure relation for $N$ Wannier excitons, which would read
\begin{equation}
\label{eq_closure_B_Wannier}
\clos{B}{N} = \frac{1}{N!} \, \sum \, \cbB{j_1} \cdots \cbB{j_N} \ket{v} \bra{v} \dbB{j_N} \cdots \dbB{j_1}
\end{equation}
for bosonized excitons~\cite{Fetter_Walecka_book_71,Martin_Rothen_book_04}, i.e., excitons with creation operators assumed to be such that $\com{\dbB{j'}}{\cbB{j}} = \delta_{j'\,j}$, is transformed into 
\begin{equation}
\label{eq_closure_Wannier}
\clos{X}{N} = \left(\frac{1}{N!}\right)^2 \, \sum \, \cB{j_1} \cdots \cB{j_N} \ket{v} \bra{v} \dB{j_N} \cdots \dB{j_1}
\end{equation}
when the exciton composite nature of the Wannier excitons is kept, i.e., when the commutator $\com{\dB{j'}}{\cB{j}}$ is taken not exactly equal to $\delta_{j'\,j}$. 

The huge prefactor change, from $1/N!$ to $(1/N!)^2$, has drastic consequences on all sum rules: indeed, the ones in the exact fermion subspace differ from the ones in the mapped bosonic subspace, whatever the mapping procedure is. Note that this difference which exists for $N=2$ excitons already, i.e. for a vanishingly small density, destroys the widely spread belief that bosonization should be valid in the small density limit at least.

Difference in closure relations has already been shown to be of practical use: it has allowed us to rederive~\cite{CBM_PRB05,CBMC_PRB07}, without calculation, the puzzling factor $1/2$ difference we found between the lifetime of $N$ ground state excitons and the sum of their scattering rates towards all the other $N$-exciton states, as first derived in~\cite{CBM_PRL04} through the explicit calculations of these two quantities separately. This quite fundamental factor $1/2$ difference between composite and elementary excitons implies that, if we manage to construct a bosonization procedure which gives correct scattering rates, we are going to miss the correct lifetime by a factor of $2$: this consequence of bosonization cannot be seen as a marginal effect, definitely.

The prefactor change from $1/N!$ to $(1/N!)^2$ in the closure relation of $N$ Wannier exciton states is however not generic for all composite bosons, as revealed when considering Frenkel excitons. Indeed, while Wannier exciton is constructed on delocalized electron and delocalized hole, so that the corresponding pair has two degrees of freedom, $\kvec{k}_e$ and $\kvec{k}_h$, Frenkel exciton is made of atomic excitations, its electron and its hole being on the same site. The corresponding pair thus has one degree of freedom only, the index $n$ of the atomic site which is excited. In a recent work~\cite{CP_arXiv_2_08}, we have shown that the closure relation for $N$ Frenkel excitons has the same $1/N!$ prefactor as the one for $N$ elementary bosons: this is after all reasonable since Frenkel exciton has one degree of freedom only, its wavevector $\kvec{Q}$ just as the localized electron-hole pairs on which they are constructed. On the contrary, in addition to its center-of-mass momentum $\kvec{Q}$, Wannier exciton has a second degree of freedom, namely its relative motion index $\eta$. The two degrees of freedom $(\eta_i,\kvec{Q}_i)$ of Wannier exciton $i$ are nothing but the memory of the two degrees of freedom of the free electrons and free holes, $\kvec{k}_e$ and $\kvec{k}_h$, out of which Wannier exciton is made. 

Another composite boson of physical interest is clearly the polariton~\cite{Hopfield_58}: being linear combination of one elementary boson, the photon, and one composite boson made of two fermions, the exciton, the polariton is a far more complicated particle: depending on the relative weight of the photon in the polariton, i.e., the so-called Hopfield coefficients, the polariton can go from a pure elementary boson to a two-fermion boson. The purpose of this paper is to show how this appears in the closure relation on $N$ polaritons. When compared to the closure relation of $N$ elementary bosons, $N$ Frenkel excitons or $N$ Wannier excitons, the closure relations of $N$ polaritons made of bosonized excitons, Frenkel excitons or Wannier excitons will allow us to better grasp the fundamental consequences of fermionic components in composite bosons. 

The paper is organized as follows: in Section~\ref{sec_2_ferm_pair}, we come back to the basic idea behind a pair of fermions behaving as a boson, to emphasize difference between elementary and composite bosons. We briefly rederive the closure relations of Wannier and Frenkel excitons with a special focus on the physical origin of the prefactor change. We also recall some fundamental results on polaritons. 

In Section~\ref{sec_5_clos_pol}, we mainly derive the closure relation for $N$ polaritons made of photons and Wannier excitons. A convenient way to do this is to start from the one in the uncoupled photon-exciton subspace. Even for $N=(2,3,4)$, the expressions obtained appear at first as rather complicated; in particular, in contrast with the closure relation for Wannier excitons, they involve non-diagonal terms in polariton operators. A careful study of these various terms however makes their physical origin rather clear: since polaritons can go from pure photons to pure excitons, this fundamental change has to appear in their closure relation; clearly we must find the elementary boson prefactor $1/N!$ if all polaritons are taken as pure photons, and the Wannier exciton prefactor $(1/N!)^2$ if they are taken as pure excitons, the exciton part of the polaritons appearing explicitely otherwise. We end this Section by studying the effect of the quantum particle composite nature through polaritons made of bosonized excitons. We also study the effect of fermionic components by considering polaritons made of Frenkel excitons. In these last two cases, the closure relation of $N$ polaritons is found to reduce to the one of $N$ elementary bosons. 

In the last Section, we compare all the closure relations for composite bosons we now have at hand. This leads us to conclude that it is not so much the composite boson nature of the particles, nor their possible fermionic components, that determine the form of their closure relation. Indeed, we find not only the elementary boson prefactor for very different composite bosons, but also very different prefactors with non-diagonal terms in the specific case of the closure relation for polaritons made of Wannier excitons. The totally compact although rather sophisticated form for the closure relation of polaritons made of Wannier excitons thus constitutes an important milestone in our understanding of the many-body physics of composite quantum particles.

\section{Elementary bosons, Wannier or Frenkel excitons, polaritons}
\label{sec_2_ferm_pair}

\subsection{Elementary/composite bosons}
\label{subsec_elem_comp_boson}

The creation operator of elementary composite bosons fulfill both $\com{\cbB{j'}}{\cbB{j}} = \cbB{j'} \cbB{j} -\cbB{j} \cbB{j'} = 0$ and $\com{\dbB{j'}}{\cbB{j}} = \, \delta_{j'\,j}$. If we now turn to composite particle made of linear combination of free fermion pairs 
\begin{equation}
\label{eq_cB}
\cB{j} = \sum \, \braket{n,m}{j} \, \ca{m} \cb{n} \ket{v}
\end{equation}
where $\ket{j} = \cB{j} \ket{v}$ while $\ket{m,n} = \ca{m}\cb{n}\ket{v}$, where $\ca{m}$ and  $\cb{n}$ create the two fermions out of which the composite particle is made, it is easy to check from the {\em anticommutators} of fermion operators, namely $\acom{\ca{m'}}{\ca{m}}     = \ca{m'} \ca{m} + \ca{m} \ca{m'} = 0 $ and $\acom{\cb{n'}}{\cb{n}} = 0$, that these composite particles behave as bosons with respect to the creation operator {\em commutator} $\com{\cB{j'}}{\cB{j}} = 0$. This holds both for fermions of different nature, as the electron and proton of an hydrogen atom, so that $\com{\ca{m}}{\cb{n}} = 0$, and for fermions having the same intrinsic nature, as the electron and hole of an exciton, so that $\acom{\ca{m}}{\cb{n}} = 0$, the hole being a valence electron absence.

By contrast, these composite particles differ definitely from elementary bosons through the other commutator
\begin{equation}
\label{eq_com_BcB}
\com{\dB{j'}}{\cB{j}} = \, \delta_{j'\,j} - D_{j'\,j}
\end{equation}
for $\cB{j}\ket{v}$ being system eigenstate, so that $\bra{v}\dB{j'}\cB{j}\ket{v} = \delta_{j'\,j}$, as easy to check from $\acom{\da{m'}}{\ca{m}} = \delta_{m'\,m}$  and $\acom{\db{n'}}{\cb{n}} = \delta_{n'\,n}$. Again it should be stressed that eq.(\ref{eq_com_BcB}) holds both for fermions having the same or different nature.

The whole purpose of the composite boson many-body theory we have recently constructed~\cite{CBMDub_PhysRep07}, is to deal with the ``deviation-from-boson operator'' $D_{j'\,j}$, appearing in~(\ref{eq_com_BcB}), exactly. It essentially generate the so-called dimensionless ``Pauli scatterings'' $\la{j'_1}{j'_2}{j_1}{j_2}$ for carrier exchanges in the absence of carrier interaction, through
\begin{equation}
\label{eq_com_dev_B}
\com{D_{j'_1\,j_1}}{\cB{j_2}} = \, \sum_{j'_2} \, \left[ \la{j'_1}{j'_2}{j_1}{j_2} + (j'_1 \leftrightarrow j'_2) \right] \, \cB{j'_2}
\end{equation}
An entire algebraic strategy, based on commutators like eqs.(\ref{eq_com_BcB}) and~(\ref{eq_com_dev_B}), is developed in~\cite{CBMDub_PhysRep07} and references therein.

\subsection{Wannier/Frenkel excitons}
\label{subsec_WFX}

Since these excitons are both made of fermion pairs while their closure relations are found to be different, it can be of interest to briefly recall the physical origin of this difference.

\subsubsection{Wannier excitons}

Wannier excitons exist in semiconductors having excitations well represented by delocalized electrons and delocalized holes. Direct Coulomb processes between one electron and one hole lead to correlated pairs, the excitons, which can be in bound or extended states. These excitons, eigenstates of the semiconductor hamiltonian, thus form a complete set in the one-electron hole pair subspace, their closure relation reading
\begin{equation}
\label{eq_closure_1_Wannier}
\clos{X}{1} = \, \sum_j \ket{j} \bra{j} = \, \sum_j \, \cB{j} \ket{v} \bra{v} \dB{j}
\end{equation}
with the sum taken over bound and extended states. Instead of correlated electron-hole pairs, i.e. excitons, we can also consider free electron-hole pair states $\ket{\kvec{k}_e,\kvec{k}_h} = \ca{\kvec{k}_e} \cb{\kvec{k}_h} \ket{v}$. They also form a complete set for the one electron-hole pair subspace, their closure relation reading
\begin{eqnarray}
\clos{e}{1} \clos{h}{1} &= &\sum_{\kvec{k}_e\,\kvec{k}_h} \ket{\kvec{k}_e,\kvec{k}_h} \bra{\kvec{k}_e,\kvec{k}_h} \nonumber \\
&= &\sum_{\kvec{k}_e\,\kvec{k}_h} \ca{\kvec{k}_e} \cb{\kvec{k}_h} \ket{v} \bra{v} \db{\kvec{k}_h} \da{\kvec{k}_e}  
\label{eq_closure_f1_Wannier}
\end{eqnarray}
From $\ket{j} = \clos{e}{1} \clos{h}{1} \ket{j}$, we readily find the link between Wannier exciton creation operators and electron-hole creation operators 
\begin{equation}
\label{eq_def_Bj}
\cB{j} = \sum_{\kvec{k}_e \,\kvec{k}_h} \,  \braket{\kvec{k}_e,\kvec{k}_h}{j} \, \ca{\kvec{k}_e} \cb{\kvec{k}_h}
\end{equation}

We now turn to $N$ pairs. Using the above equation, we can rewrite the sum in eq.~(\ref{eq_closure_Wannier}) as 
\begin{eqnarray}
\lefteqn{\clos{X}{N} = \left(\frac{1}{N!}\right)^2 \, \sum_{\{\kvec{k}\}} \, \left( \sum_{j_1} \braket{\kvec{k}'_{e_1},\kvec{k}'_{h_1}}{j_1} \braket{j_1}{\kvec{k}_{h_1},\kvec{k}_{e_1}} \right) \, \cdots } \nonumber \\
& &\qquad \qquad \quad \cdots \, \left( \sum_{j_N} \braket{\kvec{k}'_{e_N},\kvec{k}'_{h_N}}{j_N} \braket{j_N}{\kvec{k}_{h_N},\kvec{k}_{e_N}} \mbox{\rule[-.5cm]{0cm}{.7cm}} \right) \nonumber \\
& &\quad \times \, \ca{\kvec{k}'_{e_1}} \cb{\kvec{k}'_{h_1}} \cdots \ca{\kvec{k}'_{e_N}} \cb{\kvec{k}'_{h_N}}  \ket{v} \bra{v} \db{\kvec{k}_{h_N}} \da{\kvec{k}_{e_N}} \cdots \db{\kvec{k}_{h_1}} \da{\kvec{k}_{e_1}} \nonumber \\
\label{eq_closure_free_2}
\end{eqnarray}
Due to eq.~(\ref{eq_closure_1_Wannier}), the sum over $j_1$ reduces to $\delta_{\kvec{k}'_{e_1}\,\kvec{k}_{e_1}} \delta_{\kvec{k}'_{h_1}\,\kvec{k}_{h_1}}$ and similarly for the other sums. So that
\begin{equation}
\clos{X}{N} = \clos{e}{N} \clos{h}{N}
\label{eq_closure_XN}
\end{equation}
where $\clos{e}{N}$ is the closure relation for $N$ free electron states, namely, 
\begin{equation}
\label{eq_closure_N_electron}
\clos{e}{N} = \frac{1}{N!}  \, \sum_{\{\kvec{k}\}} \, \ca{\kvec{k}_1} \cdots \ca{\kvec{k}_N} \ket{v} \bra{v} \da{\kvec{k}_N} \cdots \da{\kvec{k}_1}
\end{equation}
and similarly for $N$ free holes. This readily shows that eq.~(\ref{eq_closure_Wannier}) is indeed a closure relation for the $N$ electron-hole pair subspace, one $1/N!$ coming from the $N$ free electrons and the other $1/N!$ coming from the $N$ free holes out of which the $N$ Wannier excitons are made.

\subsubsection{Frenkel excitons}

Frenkel excitons are rather different than Wannier excitons due to the underlying tight binding approximation on which these excitons are based. It makes their creation operator reading as
\begin{equation}
\label{eq_def_FQ}
\cF{\kvec{Q}} = \frac{1}{\sqrt{N_S}} \, \sum_{n=1}^{N_S} \, e^{i\kvec{Q}\cdot \kvec{R}_n} \, \cF{n}
\end{equation}
where $\cF{n} = \ca{n} \cb{n}$ creates one electron-hole pair on the same atomic site $\kvec{R}_n$, with $N_S$ being the number of atomic sites in the sample.

If we now consider one-electron states, their closure relation reads $\clos{e}{1} = \sum_n \, \ca{n} \ket{v} \bra{v} \da{n}$, and similarly for one-hole states reads $\clos{h}{1} = \sum_{n'} \, \cb{n'} \ket{v} \bra{v} \db{n'}$; so that the closure relation for all the one-electron-hole pair states appears as 
\begin{eqnarray}
\clos{e}{1} \clos{h}{1} &= &\sum_{n,n'} \, \ca{n} \cb{n'} \ket{v} \bra{v} \db{n'} \da{n} \nonumber \\
&= &\sum_{n} \, \cF{n} \ket{v} \bra{v} \dF{n} + \sum_{n\neq n'} \, \ca{n} \cb{n'} \ket{v} \bra{v} \db{n'} \da{n} \nonumber \\
\label{eq_closure_1_Frenkel}
\end{eqnarray}
We then note that states in which electron and hole are not on the same site have a much higher energy than states $\cF{n} \ket{v}$, due to the electrostatic energy cost to separate electron from hole in the tight-binding limit~\cite{CP_PRB08}. Consequently, the closure relation for the subspace made of the lowest energy electron-hole states can be reduced to the first term of eq.~(\ref{eq_closure_1_Frenkel}). This phase-space reduction is at the origin of the closure relation difference between Frenkel and Wannier excitons.

If we now use eq.~(\ref{eq_def_FQ}), we find due to lattice periodicity 
\begin{eqnarray}
\lefteqn{\clos{F}{1} = \sum_{\kvec{Q}} \, \cF{\kvec{Q}} \ket{v} \bra{v} \dF{\kvec{Q}} } \nonumber \\
&= &\frac{1}{N_S} \, \sum_{n \, n'} \, \left(\sum_{\kvec{Q}} e^{i\kvec{Q}\cdot (\kvec{R}_{n'}-\kvec{R}_n)}\right)  \, \cF{n'} \ket{v} \bra{v} \dF{n}  \nonumber \\
&= &\, \sum_{n} \, \cF{n} \ket{v} \bra{v} \dF{n} 
\label{eq_closure_1_Frenkel_2} 
\end{eqnarray}
Consequently, $\clos{F}{1}$ is a closure relation for the one-pair states belonging to the lowest energy subspace, i.e., the one of physical relevance.

We now turn to 2-Frenkel exciton states. The closure relation for 2-electron states reads
\begin{equation}
\label{eq_closure_2_electronF}
\clos{e}{2} = \frac{1}{2!} \, \sum \, \ca{n_1} \ca{n_2} \ket{v} \bra{v} \da{n_2} \da{n_1}
\end{equation}
By writing the one for 2-hole states as 
\begin{eqnarray}
\clos{h}{2} &= &\frac{1}{2!}  \sum \cb{n'_1} \cb{n'_2} \ket{v} \bra{v} \db{n'_2} \db{n'_1} \nonumber \\
&= &\frac{1}{2!}  \left[ \cb{n_1} \cb{n_2} \ket{v} \bra{v} \db{n_2} \db{n_1} + \cb{n_2} \cb{n_1} \ket{v} \bra{v} \db{n_1} \db{n_2} \mbox{\rule[-.5cm]{0cm}{.7cm}} \right.\nonumber \\ 
& &\left. \qquad + \sum_{n'_1,n'_2\neq n_1,n_2} \cb{n'_1} \cb{n'_2} \ket{v} \bra{v} \db{n'_2} \db{n'_1} \right]
\label{eq_closure_2_holeF}
\end{eqnarray}
the closure relation restricted to the subspace made of 2-electron-hole states with lowest energy  reads as
\begin{equation}
\label{eq_closure_2_el_holeF}
\clos{e}{2} \clos{h}{2} \simeq \frac{1}{2!} \left(\frac{1}{2!}\, 2\right) \, \sum_{n_1,n_2} \, \cF{n_1} \cF{n_2} \ket{v} \bra{v} \dF{n_2} \dF{n_1}
\end{equation}
Eq.(\ref{eq_def_FQ}) then allows to show that
\begin{eqnarray}
\lefteqn{\clos{F}{2} \, = \, \frac{1}{2!} \, \sum_{\kvec{Q}_1,\kvec{Q}_2} \, \cF{\kvec{Q}_1} \cF{\kvec{Q}_2} \ket{v} \bra{v} \dF{\kvec{Q}_2} \dF{\kvec{Q}_1} } \nonumber \\
&= &\frac{1}{2!} \, \frac{1}{N_S^2} \, \sum \, \left( \sum_{\kvec{Q}_1} e^{i\kvec{Q}_1\cdot (\kvec{R}_{n'_1}-\kvec{R}_{n_1})} \right.\nonumber \\
& &\left. \, \times \, \sum_{\kvec{Q}_2} e^{i\kvec{Q}_2\cdot (\kvec{R}_{n'_2}-\kvec{R}_{n_2})} \right) \, \cF{n'_1} \cF{n'_2} \ket{v} \bra{v} \dF{n_2} \dF{n_1}  
\label{eq_closure_2_el_holeF_2}
\end{eqnarray}
So that, due again to lattice periodicity, we find 
\begin{equation}
\clos{F}{2} = \frac{1}{2!} \, \sum \, \cF{n_1} \cF{n_2} \ket{v} \bra{v} \dF{n_2} \dF{n_1} \, \simeq \, \clos{e}{2} \clos{h}{2}
\label{eq_closure_2_el_holeF_3}
\end{equation}

The same procedure allows us to show that the closure relation for the lowest energy subspace made of states with electron and hole on the same site, reads~\cite{CP_arXiv_2_08}
\begin{equation}
\label{eq_closure_N_Frenkel}
\clos{F}{N} \, = \, \frac{1}{N!}  \, \sum_{\{\kvec{Q}\}} \, \cF{\kvec{Q}_1} \cdots \cF{\kvec{Q}_N} \ket{v} \bra{v} \dF{\kvec{Q}_N} \cdots \dF{\kvec{Q}_1}
\end{equation}
We wish to insist on the link which exists between the subspace reduction to the lowest energy states (which makes closure relations like the one of eq.~(\ref{eq_closure_2_el_holeF}) only approximate), and the fact that Frenkel excitons have a $1/N!$ prefactor in their closure relation, instead of a $(1/N!)^2$ as for Wannier excitons, these last excitons being made out of the $N$ available free electrons and $N$ available free holes. The change from $1/N!$ to $(1/N!)^2$ actually comes from the tight-binding approximation on which Frenkel excitons are based, and which dramatically reduces the number of pairs out which they are constructed.

\subsection{Polaritons}
\label{subsec_4_bas_pol}

Let us end this section by recalling some useful results on polaritons~\cite{Hopfield_58,CDupBM_EPL07}. Polaritons are the exact eigenstates of one photon coupled to one exciton. A convenient way to write the photon-semiconductor coupling $W_{ph-sc}$ is
\begin{equation}
\label{eq_Wphsc}
W_{ph-sc} \, = \, \sum_n \, \Omega_{n j} \, \calf{n} \dB{j} + h. \, c.
\end{equation}
where $\calf{n}$ creates one photon in the eigenmode $n$, namely $\ket{n} = \calf{n} \ket{v}$ while $\Omega_{n j}$ is the Rabi coupling of this photon to the exciton $j$. This form allows us to immediately see that photons are predominantly coupled to ground state excitons, their Rabi coupling being the largest, while the Rabi coupling to free electron-hole pairs is essentially constant for each allowed transition.

The closure relation for states made of linear combination of one photon and one exciton, 
\begin{equation}
\label{eq_closure_1f_polariton}
\clos{\nu}{1} + \clos{X}{1} = \sum_{n} \ket{n} \bra{n} + \sum_{j} \ket{j} \bra{j} 
\end{equation}
leads to write one-polariton states as $\ket{p} = (\clos{\nu}{1} + \clos{X}{1}) \ket{p}$. This shows that the polariton creation operator defined by $\cP{p} \ket{v} = \ket{p}$, reads as linear combination of photon and exciton creation operators
\begin{equation}
\label{eq_def_cP}
\cP{p} \, = \, \sum_{n} \, \braket{n}{p} \, \calf{n} + \sum_{j} \braket{j}{p} \, \cB{j} 
\end{equation}

In the same way, by using closure for one-polariton states, which are eigenstates of the coupled photon-exciton system, we have
\begin{equation}
\label{eq_closure_1_polariton_1}
\clos{P}{1} = \sum_{p} \ket{p} \bra{p} 
\end{equation}
From $\ket{n} = \clos{P}{1} \ket{n}$ and $\ket{j} = \clos{P}{1} \ket{j}$, we then find that photon and exciton creation operators can be written in terms of polaritons as
\begin{eqnarray}
\calf{n} &= & \sum_{p} \, \braket{p}{n} \, \cP{p} \label{eq_inv_def_cP_1} \\
\cB{j} &= & \sum_{p} \, \braket{p}{j} \, \cP{p}  \label{eq_inv_def_cP_2}
\end{eqnarray}
The prefactors $\braket{n}{p}$ and $\braket{j}{p}$ are nothing but the so-called Hopfield coefficients~\cite{Hopfield_58,CDupBM_EPL07}. For normalized states  $\braket{p'}{p}=\delta_{p'\,p}$ while $\braket{n'}{n} = \delta_{n'\,n}$ and $\braket{j'}{j} = \delta_{j'\,j}$, these coefficients are such that
\begin{equation}
\label{eq_norm_polariton}
\sum_{n} \norm{\braket{n}{p}}^2 + \sum_{j} \norm{\braket{j}{p}}^2 = 1
\end{equation}
This shows that polaritons can go from pure photon when $\braket{n}{p} = 1$ to pure exciton when $\braket{n}{j} = 1$, i.e., from elementary boson to 2-fermion boson. The purpose of the next section is to show how this possible change shows up in the closure relation of $N$ polaritons, starting from
\begin{equation}
\label{eq_closure_1_polariton_11}
\clos{P}{1} = \clos{\nu}{1} + \clos{X}{1} 
\end{equation}
which follows from eqs.(\ref{eq_def_cP}) and~(\ref{eq_inv_def_cP_1},\ref{eq_inv_def_cP_2}).

\section{Closure relation for polariton states}
\label{sec_5_clos_pol}

Eq.~(\ref{eq_closure_1_polariton_1}) readily gives the closure relation for $N=1$ polariton states in terms of polariton operators as
\begin{equation}
\label{eq_closure_1_polariton}
\clos{P}{1} \, = \, \sum_{p} \, \cP{p} \ket{v} \bra{v} \dP{p}
\end{equation}
Let us now see how this simple form transforms when $N$ increases. We start with polaritons made of Wannier excitons.

\subsection{$2$-polariton states}

\subsubsection{Construction of the $2$-polariton closure relation}

Due to eq.(\ref{eq_def_cP}), the two-polariton state $\cP{p_1} \cP{p_2} \ket{v}$ is a linear combination of two-photon states $\calf{n_1} \calf{n_2} \ket{v}$, two-exciton states $\cB{j_1} \cB{j_2} \ket{v}$ and one photon-one exciton states $\calf{n} \cB{j} \ket{v}$. So that the closure relation in the $2$-polariton subspace has to be the sum of the ones for these three types of states, namely
\begin{equation}
\label{eq_closure_2_polariton}
\clos{P}{2} \, = \, \clos{\nu}{2} + \clos{\nu}{1} \clos{X}{1} + \clos{X}{2} 
\end{equation}
Since photons are elementary bosons their closure relation reads as eq.(\ref{eq_closure_B_Wannier}), while in the case of Wannier excitons the closure relation is given by eq.(\ref{eq_closure_Wannier}); so that these three terms read as 
\begin{eqnarray}
\clos{\nu}{2} &= &\frac{1}{2!} \, \sum \, \calf{n_1} \calf{n_2} \ket{v} \bra{v} \dalf{n_2} \dalf{n_1} \label{eq_closure_2a_polariton} \\
\clos{\nu}{1} \clos{X}{1} &= &\frac{1}{1!} \left(\frac{1}{1!}\right)^2 \, \sum \, \calf{n_1} \cB{j_1} \ket{v} \bra{v} \dB{j_1} \dalf{n_1} \label{eq_closure_2b_polariton} \\
\clos{X}{2} &= &\left(\frac{1}{2!}\right)^2 \, \sum \, \cB{j_1} \cB{j_2} \ket{v} \bra{v} \dB{j_2} \dB{j_1}  \label{eq_closure_2c_polariton}
\end{eqnarray}

To get some confidence in the fact that the sum $\clos{P}{2}$ is indeed a closure relation for two polariton subspace, it is possible to directly check that with this expression of $\clos{P}{2}$, we do have $\clos{P}{2} \, \cP{p'} \cP{p} \ket{v} = \cP{p'} \cP{p} \ket{v}$ for any $(p',p)$. This can be done by first writing $\cP{p'}$ and $\cP{p}$ in terms of photons and excitons according to eq.(\ref{eq_def_cP}). We then use the fact that photon and exciton operators act in different subspaces, while $\bra{v}\dalf{n_2}\dalf{n_1}\calf{n'}\calf{n}\ket{v} = \delta_{n_2\, n} \delta_{n_1\, n'} + \delta_{n_2\, n'} \delta_{n_1\, n}$. This readily gives
\begin{eqnarray}
\clos{\nu}{2} \, \, \cP{p'} \cP{p} \ket{v} &= &\frac{1}{2!} \, 2 \, \sum_{n_1 n_2} \, \braket{n_1}{p'} \braket{n_2}{p} \, \calf{n_1} \calf{n_2} \ket{v}    \nonumber \\ 
\label{eq_verif_closure_2_pol_1} \\
\clos{\nu}{1} \clos{X}{1} \cP{p'} \cP{p} \ket{v} &= &\sum_{n_1 j_1} \, \left[ \braket{n_1}{p'} \braket{j_1}{p} + \braket{n_1}{p} \braket{j_1}{p'} \right] \nonumber \\
& &\qquad \qquad \qquad \times \, \calf{n_1} \cB{j_1} \ket{v}      \label{eq_verif_closure_2_pol_2}
\end{eqnarray}

To calculate $\clos{X}{2} \cP{p'} \cP{p} \ket{v}$, we must remember that, due to eqs.~(\ref{eq_com_BcB},\ref{eq_com_dev_B}), the scalar product of two Wannier excitons is given by 
\begin{eqnarray}
\lefteqn{\bra{v}\dB{j_2}\dB{j_1}\cB{j'}\cB{j}\ket{v} = } \nonumber \\
& &\delta_{j_2 \,j} \delta_{j_1\, j'} + \delta_{j_2 \,j'} \delta_{j_1\, j} - \la{j_1}{j_2}{j}{j'} - \la{j_2}{j_1}{j}{j'}
\label{eq_scal_2_X}
\end{eqnarray}
while from the two ways to form two excitons out of two free electron-hole pairs, we do have
\begin{equation}
\label{eq_2_X_recompo}
\sum_{j_1 j_2} \, \la{j_1}{j_2}{j}{j'} \, \cB{j_1} \cB{j_2} = - \cB{j'} \cB{j}
\end{equation}
Using these two equations, it becomes easy to show that
\begin{equation}
\label{eq_J2_XX_app}
\clos{X}{2} \, \cP{p'} \cP{p} \ket{v} = \, \frac{1}{4} \, 4 \, \sum_{j_1 j_2} \, \braket{j_1}{p} \braket{j_2}{p}\cB{j_1} \, \cB{j_2} \ket{v} 
\end{equation}
So that, by collecting all the terms of eqs.~(\ref{eq_verif_closure_2_pol_1},\ref{eq_verif_closure_2_pol_2},\ref{eq_J2_XX_app}), we end with
\begin{eqnarray}
\lefteqn{\clos{P}{2} \, \cP{p'} \cP{p} \ket{v} = \left[ \sum_{n_1} \, \braket{n_1}{p'} \, \calf{n_1} + \sum_{j_1} \, \braket{j_1}{p'} \, \cB{j_1} \right]  }  \nonumber \\
& &\qquad \;  \times \, \left[ \sum_{n_2} \, \braket{n_2}{p} \, \calf{n_2} + \sum_{j_2} \, \braket{j_2}{p} \, \cB{j_2} \right] \ket{v} 
\label{eq_verif_closure_2_pol_3}
\end{eqnarray}
the RHS being nothing but $\cP{p'} \cP{p} \ket{v}$, due to eq.~(\ref{eq_def_cP}). 

We are left with writing the sum in eq.~(\ref{eq_closure_2_polariton}) in terms of polaritons. Eq.~(\ref{eq_inv_def_cP_2}) gives $\cB{j}$ in terms of $\cP{p}$, so that the
exciton-exciton part of $\clos{P}{2}$ readily gives
\begin{equation}
\label{eq_J2_XX}
\clos{X}{2} = \frac{1}{4} \, \sum_{_{p_1\,p_2}^{p'_1\,p'_2}} \, \pix{p'_1}{p_1} \pix{p'_2}{p_2} \, \cP{p'_1} \cP{p'_2} \ket{v} \bra{v} \dP{p_2} \dP{p_1} 
\end{equation}
where $\pix{p'}{p}$ physically corresponds to the polariton-exciton overlap 
\begin{equation}
\label{eq_pix}
\pix{p'}{p} = \bra{p'}\clos{X}{1}\ket{p}
\end{equation}
$\clos{X}{1}$ being the projector over the one exciton subspace defined in eq.~(\ref{eq_closure_1_Wannier}). In the same way, by writing $\calf{n}$ in terms of $\cP{p}$ according to eq.(\ref{eq_inv_def_cP_1}), and by noting that, due to eq.~(\ref{eq_closure_1f_polariton}), the projector over the one-photon subspace $\clos{\nu}{1}$ can be replaced by $\clos{P}{1}-\clos{X}{1}$, we find that the photon-exciton part is made of two terms 
\begin{eqnarray}
\lefteqn{\clos{\nu}{1} \clos{X}{1} = \sum_{p'_1\,p_1\,p_2} \, \pix{p'_1}{p_1} \, \cP{p'_1} \cP{p_2} \ket{v} \bra{v} \dP{p_2} \dP{p_1} }\nonumber \\
& &\quad - \sum_{{}_{p_1\,p_2}^{p'_1\,p'_2}} \, \pix{p'_1}{p_1} \pix{p'_2}{p_2} \, \cP{p'_1} \cP{p'_2} \ket{v} \bra{v} \dP{p_2} \dP{p_1} 
\label{eq_J2_phX}
\end{eqnarray}
while the photon-photon part is made of three terms
\begin{eqnarray} 
\lefteqn{\clos{\nu}{2} = \frac{1}{2} \, \sum_{p_1\,p_2} \cP{p_1} \cP{p_2} \ket{v} \bra{v} \dP{p_2} \dP{p_1} } \nonumber \\
& &- \, \frac{1}{2} \, 2 \, \sum_{p'_1\,p_1\,p_2} \, \pix{p'_1}{p_1} \, \cP{p'_1} \cP{p_2} \ket{v} \bra{v} \dP{p_2} \dP{p_1} \nonumber \\
& &+ \, \frac{1}{2} \, \sum_{_{p_1\,p_2}^{p'_1\,p'_2}} \, \pix{p'_1}{p_1} \pix{p'_2}{p_2} \, \cP{p'_1} \cP{p'_2} \ket{v} \bra{v} \dP{p_2} \dP{p_1} 
\label{eq_J2_phph}
\end{eqnarray}
By collecting these three contributions, we find that $\clos{P}{2}$, when written in terms of polariton operators, reduces to two terms only, namely
\begin{eqnarray}
\lefteqn{\clos{P}{2} = \frac{1}{2} \, \sum_{p_1\,p_2} \cP{p_1} \cP{p_2} \ket{v} \bra{v} \dP{p_2} \dP{p_1} } \nonumber \\
& &- \, \frac{1}{4} \, \sum_{_{p_1\,p_2}^{p'_1\,p'_2}} \, \pix{p'_1}{p_1} \pix{p'_2}{p_2} \, \cP{p'_1} \cP{p'_2} \ket{v} \bra{v} \dP{p_2} \dP{p_1} 
\label{eq_I2}
\end{eqnarray}
the terms with one polariton-exciton overlap $\pix{p'}{p}$ cancelling exactly.

\subsubsection{Analysis of the $2$-polariton closure relation}

Before going further, let us physically analyse in detail the $2$-polariton closure relation~(\ref{eq_I2}). If all polaritons were just photons, the projector on exciton subspace would give zero, $\clos{X}{1}\ket{p}=0$; so that the closure relation would reduce to its first term, namely, $1/2 \, \sum \, \cP{p_1} \cP{p_2} \ket{v} \bra{v} \dP{p_2} \dP{p_1}$, this term having the same $1/2$ prefactor as the one of the closure relation for two elementary bosons. In the same way, if all the polaritons were only excitons, their projections on the exciton subspace would give $\clos{X}{1}\ket{p}=\ket{p}$; so that $\pix{p'}{p}=\delta_{p'\,p}$. Their closure relation would then read $(1/2-1/4) \, \sum \, \cP{p_1} \cP{p_2} \ket{v} \bra{v} \dP{p_2} \dP{p_1}$: we recover the closure relation of two Wannier excitons, with its $1/4$ prefactor. Since polaritons are partly photon and partly exciton, the closure relation for polaritons has to include the exciton fraction of these polaritons explicitely. This appears through the matrix elements $\pix{p'}{p}$ which allow us to go from the closure relation of pure photons with its $1/2$ prefactor, to the closure relation of pure Wannier excitons, with its $1/4$ prefactor. 

It can be of interest to note that this closure relation for polaritons could as well be written in terms of projections over the photon subspace, instead of projections over the exciton subspace. Due to eq.~(\ref{eq_closure_1_polariton_11}), these two projections are linked by 
\begin{equation}
\label{eq_pin}
\pin{p'}{p} = \bra{p'}\clos{\nu}{1}\ket{p} = \delta_{p'\,p} - \pix{p'}{p} 
\end{equation}
Using this relation, we can rewrite closure relation~(\ref{eq_I2}) in terms of $\pin{p'}{p}$'s as
\begin{eqnarray} 
\lefteqn{\clos{P}{2} = \frac{1}{4} \, \sum_{p_1\,p_2} \cP{p_1} \cP{p_2} \ket{v} \bra{v} \dP{p_2} \dP{p_1} } \nonumber \\
& &+ \, \frac{1}{2} \, \sum_{p'_1\,p_1\,p_2} \, \pin{p'_1}{p_1} \, \cP{p'_1} \cP{p_2} \ket{v} \bra{v} \dP{p_2} \dP{p_1} \nonumber \\
& &- \, \frac{1}{4} \, \sum_{_{p_1\,p_2}^{p'_1\,p'_2}} \, \pin{p'_1}{p_1} \pin{p'_2}{p_2} \, \cP{p'_1} \cP{p'_2} \ket{v} \bra{v} \dP{p_2} \dP{p_1} 
\label{eq_I2nu}
\end{eqnarray}
This second form of the closure relation for $2$-polariton states turns out to be less compact than eq.~(\ref{eq_I2}) which uses $\pix{p'}{p}$, since it contains three terms instead of two. It however allows us to readily see that, if all the polaritons were pure excitons, i.e., if all the $\pin{p'}{p}$'s give zero, we get the closure relation for $2$ Wannier excitons with its $1/4$ prefactor. Note that eq.~(\ref{eq_I2nu}) also allows to recover that, if all polaritons were pure photons, i.e., if all the $\pin{p'}{p}$'s reduce to $\delta_{p'\,p}$, the closure relation reduces to $(1/4+1/2-1/4) \, \sum \, \cP{p_1} \cP{p_2} \ket{v} \bra{v} \dP{p_2} \dP{p_1}$, with the same $1/2$ prefactor as the one for two elementary bosons.

A third possibility is to use the difference between the exciton and photon weights of the polariton, namely
\begin{equation}
\label{eq_pixmn}
\pim{p'}{p} = \bra{p'}\left(\clos{X}{1}-\clos{\nu}{1}\right)\ket{p} 
\end{equation}
Eqs.~(\ref{eq_pin},\ref{eq_pixmn}) then give the closure relation for $2$-polariton states, eq.~(\ref{eq_I2}), as
\begin{eqnarray}
\lefteqn{\clos{P}{2} = \frac{7}{16} \, \sum_{p_1\,p_2} \cP{p_1} \cP{p_2} \ket{v} \bra{v} \dP{p_2} \dP{p_1} } \nonumber \\
& &- \, \frac{1}{8} \, \sum_{p'_1\,p_1\,p_2} \, \pim{p'_1}{p_1} \, \cP{p'_1} \cP{p_2} \ket{v} \bra{v} \dP{p_2} \dP{p_1} \nonumber \\
& &- \, \frac{1}{16} \, \sum_{_{p_1\,p_2}^{p'_1\,p'_2}} \, \pim{p'_1}{p_1} \pim{p'_2}{p_2} \, \cP{p'_1} \cP{p'_2} \ket{v} \bra{v} \dP{p_2} \dP{p_1} 
\label{eq_I2diff}
\end{eqnarray}
In the case of 2D-microcavity polaritons, these equations could then be further simplified in terms of the explicit Hopfield coefficient for the upper and lower polariton branches, as done in ref.~\cite{CDupBM_EPL07}.

\subsection{$3$- and $4$-polariton states} 

According to eq.~(\ref{eq_def_cP}), the $3$-polariton state $\cP{p''}\cP{p'}\cP{p_2}\ket{v}$ contains states with three photons, states with two photons and one exciton, states with one photon and two excitons, and states with three excitons. So that the sum
\begin{equation}
\clos{P}{3} = \clos{\nu}{3} + \clos{\nu}{2} \clos{X}{1} + \clos{\nu}{1} \clos{X}{2} + \clos{X}{3}
\label{eq_I3_pol}
\end{equation}
is a closure relation for $3$-polariton states.

Eqs.~(\ref{eq_closure_Wannier}) and~(\ref{eq_inv_def_cP_2}) readily give the closure relation for states made of $3$ excitons as
\begin{eqnarray}
\clos{X}{3} &= &\left(\frac{1}{3!}\right)^2 \, \sum_{\{j\}} \, \cB{j_1} \cB{j_2} \cB{j_3} \ket{v} \bra{v} \dB{j_3} \dB{j_2} \dB{j_1} \nonumber \\
&= &\left(\frac{1}{3!}\right)^2 \, \sum_{\{p\}} \, \pix{p'_1}{p_1} \pix{p'_2}{p_2} \pix{p'_3}{p_3} \nonumber \\
& &\qquad \qquad \times \, \cP{p'_1} \cP{p'_2} \cP{p'_3} \ket{v} \bra{v} \dP{p_3} \dP{p_2} \dP{p_1} 
\label{eq_I3_XXX}
\end{eqnarray}
If we now turn to the closure relation in the subspace made of $3$-photon states, eq.~(\ref{eq_inv_def_cP_1}) allows us to write it as
\begin{eqnarray}
\clos{\nu}{3} &= &\frac{1}{3!} \, \sum_{\{n\}} \, \calf{n_1} \calf{n_2} \calf{n_3} \ket{v} \bra{v} \dalf{n_3} \dalf{n_2} \dalf{n_1} \nonumber \\
&= &\frac{1}{3!} \, \sum_{\{p\}} \, \pin{p'_1}{p_1} \pin{p'_2}{p_2} \pin{p'_3}{p_3} \nonumber \\
& &\qquad \qquad \times \, \cP{p'_1} \cP{p'_2} \cP{p'_3} \ket{v} \bra{v} \dP{p_3} \dP{p_2} \dP{p_1} 
\label{eq_I3_phphph}
\end{eqnarray}
By using eq.~(\ref{eq_pin}) to write $\pin{p'}{p}$ in terms of $\pix{p'}{p}$, and by relabelling the bold indices, this $\clos{\nu}{3}$ operator can be rewritten as a sum of terms with zero, one, two and three $\pix{p'}{p}$: 
\begin{eqnarray}
\clos{\nu}{3} &= &\frac{1}{3!} \, \sum \, \cP{p_1} \cP{p_2} \cP{p_3} \ket{v} \bra{v} \dP{p_3} \dP{p_2} \dP{p_1}  \nonumber \\
& &- \, \frac{1}{3!} \, 3 \, \sum \, \pix{p'_3}{p_3} \, \cP{p_1} \cP{p_2} \cP{p'_3} \ket{v} \bra{v} \dP{p_3} \dP{p_2} \dP{p_1} \nonumber \\
& &+ \, \frac{1}{3!} \, 3 \, \sum \, \pix{p'_2}{p_2} \pix{p'_3}{p_3} \nonumber \\
& &\qquad \qquad \quad \times \, \cP{p_1} \cP{p'_2} \cP{p'_3} \ket{v} \bra{v} \dP{p_3} \dP{p_2} \dP{p_1} \nonumber \\
& &- \, \frac{1}{3!} \, \sum \, \pix{p'_1}{p_1} \pix{p'_2}{p_2} \pix{p'_3}{p_3} \nonumber \\
& &\qquad \qquad \quad \times \, \cP{p'_1} \cP{p'_2} \cP{p'_3} \ket{v} \bra{v} \dP{p_3} \dP{p_2} \dP{p_1} 
\label{eq_I3_phphph_2}
\end{eqnarray}
If we now perform the same kind of transformations for the closure relation of states made of two photons and one exciton, namely,
\begin{equation}
\label{eq_I3_phphX}
\clos{\nu}{2} \clos{X}{1} = \, \frac{1}{2!} \left(\frac{1}{1!}\right)^2 \, \sum \, \calf{n_1} \calf{n_2} \cB{j_1} \ket{v} \bra{v} \dB{j_1} \dalf{n_2} \dalf{n_1}
\end{equation}
and for states made of one photon and two excitons, namely,
\begin{equation}
\label{eq_I3_phXX}
\clos{\nu}{1} \clos{X}{2} = \, \frac{1}{1!} \left(\frac{1}{2!}\right)^2 \, \sum \, \calf{n_1} \cB{j_1} \cB{j_2} \ket{v} \bra{v} \dB{j_2} \dB{j_1} \dalf{n_1}
\end{equation}
we find that the closure relation for $3$ polaritons which correspond to states made of $3$ photons, $2$ photons and $1$ exciton, $1$ photon and $2$ excitons, and $3$ excitons, reads in terms of polariton operators as
\begin{eqnarray}
\clos{P}{3} &= &\frac{1}{3!} \, \sum \, \cP{p_1} \cP{p_2} \cP{p_3} \ket{v} \bra{v} \dP{p_3} \dP{p_2} \dP{p_1}  \nonumber \\
& &- \, \frac{1}{4} \, \sum \, \pix{p'_1}{p_1} \pix{p'_2}{p_2} \, \cP{p'_1} \cP{p'_2} \cP{p_3} \ket{v} \bra{v} \dP{p_3} \dP{p_2} \dP{p_1} \nonumber \\
& &+ \, \frac{1}{9} \, \sum \, \pix{p'_1}{p_1} \pix{p'_2}{p_2} \pix{p'_3}{p_3} \, \nonumber \\
& &\qquad \qquad \quad \times \, \cP{p'_1} \cP{p'_2} \cP{p'_3} \ket{v} \bra{v} \dP{p_3} \dP{p_2} \dP{p_1} 
\label{eq_I3}
\end{eqnarray}

We note that, as for $\clos{P}{2}$, given in eq.~(\ref{eq_I2}), $\clos{P}{3}$ does not contain term linear in polariton-exciton overlap $\pix{p'}{p}$. We also note that the sum of the three prefactors, namely $(1/3!-1/4+1/9)$ is equal to $(1/3!)^2$, as for $3$ Wannier excitons. Indeed, if all polaritons were pure excitons, $\clos{X}{1}\ket{p}$ would be equal to $\ket{p}$ for all $p$, so that we should find the closure relation for $3$-exciton states with its $(1/3!)^2$ prefactor. In the same way, we see that, if all polaritons were pure photons, their projection $\clos{X}{1}\ket{p}$ on the exciton subspace would reduce to zero; so that $\clos{P}{3}$ would reduce to its first term - which is nothing but the closure relation of $3$ photons, i.e., $3$ elementary bosons. Contributions to the closure relation of polaritons which are partly photon and partly exciton are far more complicated; they have to contain the excitonic weight of these polaritons explicitely, as seen from eq.~(\ref{eq_I3}).

If we now consider $4$-polariton states, and follow a similar procedure, we find a term with no polariton-exciton overlap $\pix{p'}{p}$, its prefactor being $(1/4!)$, no term with one $\pix{p'}{p}$, a term with two $\pix{p'}{p}$, its prefactor being $(-1/8)$, a term with three $\pix{p'}{p}$, its prefactor being $1/9$, and a term with four $\pix{p'}{p}$, its prefactor being $-15/(4!)^2$. Although these prefactors are individually not physically meaningful, their sum $[1/4!-1/8+1/9-15/(4!)^2]$ reduces to $(1/4!)^2$ as expected for $4$ Wannier excitons: here again, we go from prefactor $(1/4!)$ for $4$ polaritons being pure photons, as obtained with the term having zero $\pix{p'}{p}$, to $(1/4!)^2$ for $4$ polaritons being pure Wannier excitons, when all the $\pix{p'}{p}$ are equal to $1$.

\subsection{$N$-polariton states made of Wannier excitons}

We keep using the same procedure and start from the sum of closure relations in the subspace made of $N$ photons, $(N-1)$ photons and one exciton, $(N-2)$ photons and two excitons, and so on... up to the subspace made of $N$ excitons solely. This leads to 
\begin{equation}
\label{eq_closure_IN_sum_kph}
\clos{P}{N}  = \sum_{k=0}^N \clos{\nu}{k} \clos{X}{N-k} 
\end{equation} 
According to eqs.~(\ref{eq_closure_B_Wannier},\ref{eq_closure_Wannier}), we can rewrite the closure relation products as 
\begin{equation}
\label{eq_closure_kph-NkX}
\clos{\nu}{k} \clos{X}{N-k} = \frac{1}{k!} \,\left(\frac{1}{(N-k)!}\right)^2 \, \snk{N}{k} 
\end{equation}
in which we have set
\begin{eqnarray}
\snk{N}{k} &= &\sum_{\{i\}\,\{j\}} \, \calf{i_1} \cdots \calf{i_{k}} \cB{j_{k+1}} \cdots \cB{j_{N}} \ket{v} \nonumber \\
& &\qquad \, \times \, \bra{v} \dB{j_{N}} \cdots \dB{j_{k+1}} \dalf{i_{k}} \cdots \dalf{i_1}
\label{eq_SNk}
\end{eqnarray}
We then use eqs.~(\ref{eq_inv_def_cP_1},\ref{eq_inv_def_cP_2}) to rewrite photon and exciton operators in terms of polaritons. This leads to
\begin{eqnarray}
\snk{N}{k} &= &\sum_{\{p'\}\,\{p\}} \, \left(\prod_{i=1}^{k} \pin{p'_i}{p_i} \right) \, \left(\prod_{j=k+1}^{N} \pix{p'_j}{p_j} \right) \nonumber \\
& &\qquad \times \, \cP{p'_1} \cdots \cP{p'_{N}} \ket{v} \bra{v} \dP{p_{N}} \cdots \dP{p_1} 
\label{eq_sum_kph-NkX_1}
\end{eqnarray}
We can now choose to only keep $\pix{p'}{p}$, or $\pin{p'}{p}$, or to introduce their difference $\pim{p'}{p}$.

\subsubsection{$N$-polariton closure relation in terms of exciton weights}

To get the closure relation for $N$ polaritons in terms of exciton weights, we rewrite $\pin{p'}{p}$ as $\delta_{p'\,p} - \pix{p'}{p}$ using eq.~(\ref{eq_pin}). By relabelling the bold indices, we end with a closure relation for $N$ polariton states which has the same structure as the one previously found, with terms a priori having $(0,1,2,\ldots ,N)$ polariton-exciton overlaps $\pix{p'}{p}$. 
\begin{eqnarray}
\clos{P}{N}  &= &\frac{1}{N!} \, \sum \, \cP{p_1} \cdots \cP{p_N} \ket{v} \bra{v} \dP{p_N} \cdots \dP{p_1}  \nonumber \\
& &+ \, \gamn{1} \, \sum \, \pix{p'_1}{p_1} \nonumber \\
& &\qquad \quad \times \, \cP{p'_1} \cP{p_2} \cdots \cP{p_N} \ket{v} \bra{v} \dP{p_N} \cdots \dP{p_2} \dP{p_1} \nonumber \\
& &+ \, \gamn{2} \, \sum \, \pix{p'_1}{p_1} \pix{p'_2}{p_2} \nonumber \\
& &\quad \times \, \cP{p'_1} \cP{p'_2} \cP{p_3} \cdots\cP{p_N} \ket{v} \bra{v} \dP{p_N} \cdots \dP{p_3} \dP{p_2} \dP{p_1} \nonumber \\
& &+ \, \cdots \nonumber \\
& &+ \, \gamn{N} \, \sum \, \pix{p'_1}{p_1} \pix{p'_2}{p_2} \cdots \pix{p'_N}{p_N} \nonumber \\
& &\qquad \qquad \times \, \cP{p'_1} \cdots \cP{p'_N} \ket{v} \bra{v} \dP{p_N} \cdots \dP{p_1} 
\label{eq_IN}
\end{eqnarray}
The above equation already shows one important feature of the closure relation for $N$ polaritons made of Wannier excitons, compared to the photon one written in eq.~(\ref{eq_closure_B_Wannier}) and the Wannier exciton one written in eq.(\ref{eq_closure_Wannier}): $\clos{P}{N}$, except for its first term, is not diagonal with respect to polariton operators, the term with $n$ overlaps having $n$ polaritons changing from state $p$ to state $p'$.

The prefactors $\gamn{n}$ in this sum follow from the prefactors of closure relations in the independent photon-exciton subspaces and the various relabelling of bold indices. The prefactor of the term with one $\pix{p'}{p}$ is found to reduce to zero whatever $N$ is, in agreement with what we previously found, since its general form reads
\begin{equation}
\label{eq_gamn1}
\gamn{1} \, = \, - \, \frac{\comb{N}{1}}{N!} + \frac{\comb{N-1}{0}}{(N-1)!} \left(\frac{1}{1!}\right)^2 \, = \, 0
\end{equation}
where $\comb{N}{n} = N!/n!(N-n)!$ is the number of ways to choose $n$ overlaps among $N$. If we now consider the prefactor of the term with two $\pix{p'}{p}$, it reads
\begin{eqnarray}
\gamn{2} &= &\frac{\comb{N}{2}}{N!} - \frac{\comb{N-1}{1}}{(N-1)!} \left(\frac{1}{1!}\right)^2 + \frac{\comb{N-2}{0}}{(N-2)!} \left(\frac{1}{2!}\right)^2 \nonumber \\
&= & - \, \frac{1}{2^2(N-2)!}
\label{eq_gamn2}
\end{eqnarray}
In the same way, the prefactor of the term with three $\pix{p'}{p}$, reads
\begin{eqnarray}
\gamn{3} &= &- \, \frac{\comb{N}{3}}{N!} + \frac{\comb{N-1}{2}}{(N-1)!} \left(\frac{1}{1!}\right)^2\nonumber \\
& &\quad - \frac{\comb{N-2}{1}}{(N-2)!} \left(\frac{1}{2!}\right)^2 + \frac{\comb{N-3}{0}}{(N-3)!} \left(\frac{1}{3!}\right)^2 \nonumber \\
&= &- \, \frac{1}{3^2} \, \frac{1}{(N-3)!}
\label{eq_gamn3}
\end{eqnarray}
and the prefactor of the term with four $\pix{p'}{p}$ is found to be
\begin{eqnarray}
\gamn{4} &= & \frac{\comb{N}{4}}{N!} - \frac{\comb{N-1}{3}}{(N-1)!} \left(\frac{1}{1!}\right)^2 + \frac{\comb{N-2}{2}}{(N-2)!} \left(\frac{1}{2!}\right)^2 \nonumber \\
& &\quad - \frac{\comb{N-3}{1}}{(N-3)!} \left(\frac{1}{3!}\right)^2 + \frac{\comb{N-4}{0}}{(N-4)!} \left(\frac{1}{4!}\right)^2 \nonumber \\
&= &- \, \frac{15}{(4!)^2} \, \frac{1}{(N-4)!}
\label{eq_gamn4}
\end{eqnarray}
The above results agree with the ones we previously found. While $\gamn{2}$ and $\gamn{3}$ would lead to guess a raher simple form for $\gamn{n}$, its value with $n=4$ actually shows that this is not so. It is however possible to put $\gamn{n}$ in a compact form in the following way. For general $n$, the prefactor $\gamn{n}$ of the terms with $n$ overlaps $\pix{p'}{p}$ reads, due to the various prefactors of the closure relations for photons and excitons, and the possible relabelling of bold indices, as
\begin{equation}
\gamn{n} = \sum_{p=0}^{n} (-1)^{n-p} \,\comb{N-p}{n-p}\, \left[ \frac{1}{(N-p)!} \left(\frac{1}{p!}\right)^2 \right]
\label{eq_gamnN_series1}
\end{equation}
By writing $\comb{N}{n}$ as $N!/n!(N-n)!$, this leads to
\begin{eqnarray}
\gamn{n} &= &\frac{(-1)^n}{(N-n)!} \, \sum_{p=0}^{n} \, \frac{(-1)^p}{(n-p)!} \left(\frac{1}{p!}\right)^2 \nonumber \\
&= &\frac{(-1)^n}{(N-n)!\,n!} \, \hypergF{-n}{1}{1}
\label{eq_gamnN_series2}
\end{eqnarray}
where $\hypergF{\alpha}{\gamma}{z}$ is the degenerate hypergeometric function (or confluent hypergeometric function of the first kind~\cite{AbramSteg,GradshRyz}), defined as
\begin{equation}
\hypergF{\alpha}{\gamma}{z} = 1 + \frac{\alpha}{\gamma} \frac{z}{1!} + \frac{\alpha (\alpha + 1)}{\gamma (\gamma + 1)} \frac{z^2}{2!} + \cdots
\label{eq_hyperG_F}
\end{equation}
Since $\hypergF{0}{1}{1}=1$, we readily find $\gamn{0}=1/N!$. In the same way, $\hypergF{-1}{1}{1} =0$ leads to $\gamn{1}= 0$, while $\hypergF{-2}{1}{1} = -1/2$ leads to $\gamn{1} = -1/(4(N-2)!)$; and so on~\ldots

Eq.~(\ref{eq_IN}) shows that, if all the polaritons were pure photons, i.e. if all the $\pix{p'}{p}$'s are equal to $0$, the closure relation would have the prefactor $1/N!$ of $N$ elementary bosons. In order to show that the closure relation $\clos{P}{N}$ reduces to the one of $N$ Wannier excitons, with its prefactor $(1/N!)^2$, if all the polaritons were pure excitons, i.e. if all the $\pix{p'}{p}$'s are equal to $1$, we must show, as explicitely checked for $N=(2,3,4)$, that the sum of all prefactors appearing in eq.~(\ref{eq_IN}) reduces to $(1/N!)^2$, namely
\begin{equation}
\label{eq_sum_prefactors}
\frac{1}{N!} + \sum_{n=1}^N \gamn{n} \, = \, \left(\frac{1}{N!}\right)^2
\end{equation}
We fist note that, using eq.~(\ref{eq_gamnN_series2}), this equation also reads $Z_N=1/N!$ where 
\begin{equation}
\label{eq_ZN}
Z_N \, = \, \sum_{n=0}^N (-1)^n \, \comb{N}{n} \, \sum_{p=0}^n (-1)^p \, \frac{\comb{n}{p}}{p!}
\end{equation}
To show that $Z_N=1/N!$, we first note that $\sum_{n=0}^N \sum_{p=0}^n = \sum_{p=0}^N \sum_{n=p}^N$ while $\comb{N}{n} \, \comb{n}{p} = \comb{N}{p} \, \comb{N-p}{n-p}$ which follows from the definition of $\comb{N}{n}$. This allows us to rewrite $Z_N$ as
\begin{equation}
\label{eq_ZN2}
Z_N \, = \, \sum_{p=0}^N \frac{(-1)^p}{p!} \, \comb{N}{p} \, \sum_{n=p}^N (-1)^n \, \comb{N-p}{n-p}
\end{equation}
To go further, we set $n'=n-p$ in the last term and note that
\begin{equation}
(1-x)^{(N-p)} \, = \, \sum_{n'=0}^{N-p} (-1)^{n'} \, \comb{N-p}{n'} \, x^{n'} \nonumber
\end{equation}
For $x=1$ this shows that the sum over $n$ in eq.~(\ref{eq_ZN2}) reduces to $\delta_{N\,p}$. So that eq.~(\ref{eq_ZN2}) leads to
\begin{equation}
\label{eq_ZN3}
Z_N \, = \, \frac{\comb{N}{N}}{N!} \, = \, \frac{1}{N!}
\end{equation}
as necessary to prove eq.~(\ref{eq_sum_prefactors}). Hence we recover the closure relation for Wannier exciton~(\ref{eq_closure_Wannier}) if all the polaritons were pure excitons.

We thus conclude that, although the closure relation for $N$ polariton states written in eq.~(\ref{eq_IN}) is definitely far more complicated than the one for $N$ elementary bosons or for $N$ Wannier excitons, it is yet compact. From it, we can recover expressions like eqs.~(\ref{eq_closure_B_Wannier}) and~(\ref{eq_closure_Wannier}), with prefactors $1/N!$  and $(1/N!)^2$ when considering polaritons as pure photons or as pure Wannier excitons. However, since polaritons have a mixed nature, as seen through the projection $\clos{X}{1}\ket{p}$ which for most polaritons differs from $0$ and $\ket{p}$, the general form of the closure relation for $N$ polariton states~(\ref{eq_IN}) has to be more complicated than the one of $N$ photons or $N$ Wannier excitons; in particular most of its terms are non-diagonal with respect to polariton operators. 

\subsubsection{$N$-polariton closure relation in terms of photon weights}

To get the closure relation for $N$ polaritons in terms of photon weights, we just have to rewrite in eq.~(\ref{eq_sum_kph-NkX_1}) $\pix{p'}{p}$ as $\delta_{p'\,p} - \pin{p'}{p}$ in order to only keep projectors $\clos{\nu}{1}$ over the photon subspace. The closure relation for $N$-polariton states then appears as
\begin{eqnarray}
\clos{P}{N} &= &\sum_{n=0}^N \, \tgamn{n} \, \sum_{\{p'\}\,\{p\}} \, \left(\prod_{i=1}^{N-n} \pin{p'_i}{p_i} \right) \nonumber \\
& &\times \cP{p'_1} \cdots \cP{p'_{n}} \cP{p_{n+1}} \cdots \cP{p_{N}} \ket{v} \bra{v} \dP{p_{N}} \cdots \dP{p_1} \nonumber \\
\label{eq_IN_ph}
\end{eqnarray}
Prefactors $\tgamn{n}$ are similar to prefactors $\gamn{n}$ in the closure relation~(\ref{eq_IN}) when written in terms of exciton weights, except that $p!$ and $(N-p)!$ are exchanged, as reasonable due to their physical origin. They precisely read
\begin{equation}
\tgamn{n} = \sum_{p=0}^{n} \, (-1)^{n-p} \, \comb{N-p}{n-p} \, \frac{1}{p!} \,\left(\frac{1}{(N-p)!}\right)^2 \, 
\label{eq_tgamnN_series}
\end{equation}
This immediately shows that $\tgamn{0} = (1/N!)^2$; so that eq.~(\ref{eq_tgamnN_series}) readily gives a closure relation with the Wannier exciton prefactor if all polaritons were excitons, i.e., if all $\pin{p'}{p}$ reduce to zero. 

As for $\gamn{n}$, it is actually possible to write $\tgamn{n}$ in a similar way in terms of hypergeometric functions, according to
\begin{eqnarray}
\tgamn{n} &= &\frac{(-1)^n}{(N-n)!} \,\sum_{p=0}^{n} \, (-1)^{p} \, \frac{1}{p!(N-p)!(n-p)!} \nonumber \\
&= &\frac{(-1)^n}{N! \,(N-n)! \,n!} \, \, U(-n, 1 - n + N; 1) \nonumber \\
&= &\frac{(-1)^n}{N! \,(N-n)! \,n!} \, \, G(-n, - N; -1)
\label{eq_tgamnN_series_2}
\end{eqnarray}
where the hypergeometric function $U(a,c;z)$, also called confluent hypergeometric function of the second kind~\cite{AbramSteg,GradshRyz}, is linked to the function $\hypergG{\alpha}{\beta}{z}$ introduced by Landau~\cite{LandauLifMQ}
\begin{equation}
\hypergG{\alpha}{\beta}{z} = 1 + \frac{\alpha \, \beta}{1!\,z} + \frac{\alpha (\alpha + 1)\, \beta(\beta + 1)}{2!\,z^2} + \cdots
\label{eq_hyperG_G}
\end{equation} 
through $U(a,c;z) = z^{-a} \, \hypergG{a}{a+1-c}{-z}$. 

In the same way as for eq.~(\ref{eq_sum_prefactors}), it is possible to show that 
\begin{equation}
\label{eq_sum_prefactors_gg}
\left(\frac{1}{N!}\right)^2 + \sum_{n=1}^N \tgamn{n} \, = \, \frac{1}{N!}
\end{equation}
in order to go from the $(1/N!)^2$ prefactor of Wannier excitons to the $1/N!$ prefactor of elementary bosons.

\subsubsection{$N$-polariton closure relation in terms of weight differences}

The closure relation for $N$ polaritons in terms of weight differences, which is an alternative to eq.~(\ref{eq_IN}) and~(\ref{eq_IN_ph}), can be derived by using the weight difference $\pim{p'}{p}$ defined in eq.~(\ref{eq_pixmn}). This projector difference is quite interesting since it treats on an equal footing the excitonic weight and the photonic weight of the polariton; therefore, we expect prefactors equivalent to $\gamn{n}$ or $\tgamn{n}$, in eqs.~(\ref{eq_IN}) or~(\ref{eq_IN_ph}), to be somewhat intermediate between $1/N!$ and $(1/N!)^2$. 

By writing $\pin{p'}{p}$ as $(\delta_{p'\,p} - \pim{p'}{p})/2$ and $\pix{p'}{p}$ as $(\delta_{p'\,p} + \pim{p'}{p})/2$, we see that, in eq.~(\ref{eq_sum_kph-NkX_1}), the first product on $i$ (respectively the second product on $j$), contains $\comb{k}{l}$ (respectively $\comb{N-k}{m}$) terms that appear with exactly $l$ (respectively $m$) factors of the type $-\pim{p'}{p}$ (respectively $+\pim{p'}{p}$), the other products involving trivial $\delta_{p'\,p}$. By relabelling the $\{p'\}$ and $\{p\}$ indices in every term of sum~(\ref{eq_sum_kph-NkX_1}), we find, after commutations of the corresponding $\cP{p'_j}$ and $\dP{p_j}$ operators, that {\em all these terms look identical and have a $\delta_{p'\,p}$ for the last $(l+m)$ indices}. Using them to simplify the sums, we get
\begin{eqnarray}
\snk{N}{k} &= &\frac{1}{2^N} \, \sum_{l=0}^{k} \, \sum_{m=0}^{N-k} \, (-1)^{l} \,\, \comb{k}{l} \,\, \comb{N-k}{m} \nonumber \\
& &\times \, \sum_{\{p'\}\,\{p\}} \, \left(\prod_{i=1}^{l+m} \pim{p'_i}{p_i} \right) \nonumber \\
& &\times \cP{p'_1} \cdots \cP{p'_{l+m}} \cP{p_{l+m+1}} \cdots \cP{p_{N}} \ket{v} \bra{v} \dP{p_{N}} \cdots \dP{p_1}  \nonumber \\
\label{eq_sum_kph-NkX_3}
\end{eqnarray}
We now come back to $\clos{P}{N}$ defined in eqs.~(\ref{eq_closure_IN_sum_kph},\ref{eq_closure_kph-NkX}), and substitute eq.~(\ref{eq_sum_kph-NkX_3}). By interchanging summations $\sum_{k=0}^N \sum_{l=0}^{k} \sum_{m=0}^{N-k}$ as $\sum_{l=0}^N \sum_{m=0}^{N-l} \sum_{k=l}^{N-m}$, we can group all terms in the following way 
\begin{eqnarray}
\lefteqn{ \clos{P}{N} = \frac{1}{2^N} \, \sum_{l=0}^N \sum_{m=0}^{N-l} \sum_{k=l}^{N-m} \, \frac{(-1)^{l}}{l!(k-l)! \, (N-k)! \, m!(N-k-m)!}  } \nonumber \\
& &\times \, \sum_{\{p'\}\,\{p\}} \, \left(\prod_{i=1}^{l+m} \pim{p'_i}{p_i}  \right) \nonumber \\
& &\times \, \cP{p'_1} \cdots \cP{p'_{l+m}} \cP{p_{l+m+1}} \cdots \cP{p_{N}} \ket{v} \bra{v} \dP{p_{N}} \cdots \dP{p_1} \nonumber \\
\label{eq_IN_diff_2}
\end{eqnarray}
By introducing $n \equiv (l+m)$, we finally get after simplification of all indices
\begin{eqnarray}
\clos{P}{N} &= &\sum_{n=0}^N \, \ggamn{n} \, \sum_{\{p'\}\,\{p\}} \, \left(\prod_{i=1}^{n} \pim{p'_i}{p_i} \right) \nonumber \\
& &\times \cP{p'_1} \cdots \cP{p'_{n}} \cP{p_{n+1}} \cdots \cP{p_{N}} \ket{v} \bra{v} \dP{p_{N}} \cdots \dP{p_1} \nonumber \\
\label{eq_IN_diff}
\end{eqnarray}
which is an alternative form for the closure relation given in eqs.(\ref{eq_IN}) and~(\ref{eq_IN_ph}), but written in terms of weight differences $\pim{p'}{p}$.
$\ggamn{n}$ in this sum appear as the following series
\begin{equation}
\ggamn{n} = \frac{(-1)^{n}}{2^N \,N!} \,\comb{N}{n} \sum_{l=0}^{n} \, \frac{(-1)^{l}}{l!} \,\, \comb{n}{l} \, \hypergF{n-N}{1+l}{-1}
\label{eq_ggamnN_series}
\end{equation}
of degenerate hypergeometric function $\hypergF{\alpha}{\gamma}{z}$ defined in  eq.~(\ref{eq_hyperG_F}). Unfortunately, we could not find for them expressions as compact as eq.(\ref{eq_gamnN_series2}) for $\gamn{n}$, or eq.(\ref{eq_tgamnN_series_2}) for $\tgamn{n}$. We can however check that, for $N=2$, eq.~(\ref{eq_ggamnN_series}) leads to the factors $7/16,-1/8,-1/16$ obtained in eq.~(\ref{eq_I2diff}). This new form of closure relation is the relevant one when the polaritons of interest are half photons/half excitons, as in the case of 2D microcavity polaritons close to zone-center when the photon detuning is very small (c.f. ref.~\cite{CDupBM_EPL07}).

\subsection{Effects of fermionic components in polaritons}
\label{subsec_6_underl_ferm}

In order to better grasp the importance of fermionic components in the closure relation of composite bosons, let us add two more closure relations to the four we already have at hand, namely the ones for bozonized excitons, Wannier excitons, Frenkel excitons and polaritons made of Wannier excitons. These are the ones for polaritons made of bosonized excitons and polaritons made of Frenkel excitons. We are going to show that, in these two cases, in spite of the composite nature of the quantum particles, the closure relation is just the one of $N$ elementary bosons.

In order to make things simpler, we restrict the derivation of these new closure relations to the case of $2$-polariton states. The closure relation $\clos{\bar{P}}{2}$ for two polaritons $\bar{P}$ either made of Frenkel excitons, or made of bosonized excitons, reads as $\clos{\nu}{2}+\clos{\nu}{1} \clos{\bar{B}}{1}+\clos{\bar{B}}{2}$. This leads to 
\begin{eqnarray}
\clos{\bar{P}}{2}  &= &\frac{1}{2!} \, \sum \, \calf{n_1} \calf{n_2} \ket{v} \bra{v} \dalf{n_2} \dalf{n_1} \nonumber \\
& &+ \, \sum \, \calf{n_1} \cbB{j_1} \ket{v} \bra{v} \dbB{j_1} \dalf{n_1} \nonumber \\
& &+ \, \frac{1}{2!} \, \sum \, \cbB{j_1} \cbB{j_2} \ket{v} \bra{v} \dbB{j_2} \dbB{j_1} 
\label{eq_closure_2_B_polariton}
\end{eqnarray}
Let us stress that, when compared to eq.~(\ref{eq_closure_2_polariton}), the prefactor of the last term is $1/2!$ instead of $(1/2!)^2$, due to the closure relation of the excitons on which these polaritons are constructed. If we now write these photon and exciton operators in terms of polaritons by using the closure relation for one polariton states, namely $\clos{\bar{P}}{1} = \sum \ket{\bar{p}}\bra{\bar{p}}$, we find that $\clos{\bar{P}}{2}$ also reads
\begin{eqnarray}
\clos{\bar{P}}{2} &= &\frac{1}{2!} \, \sum \, \left( \pin{p'_1}{p_1} \pin{p'_2}{p_2} + 2 \, \pin{p'_1}{p_1} \pix{p'_2}{p_2}  \right.  \nonumber \\
& & \quad  \left.+ \,\pix{p'_1}{p_1} \pix{p'_2}{p_2} \right) \, \cbP{p'_1} \cbP{p'_2} \ket{v} \bra{v} \dbP{p_2} \dbP{p_1}  
\label{eq_closure_2_B_polariton_2}
\end{eqnarray}
After relabelling $(p'_2,p_2)$ into $(p'_1,p_1)$ in half of the second term, the factor in parenthesis reveals products of $(\pin{p'}{p}+\pix{p'}{p})$ which all reduce to $\braket{p'}{p}=\delta_{p'\,p}$. So that the closure relation for two such polariton states
\begin{equation}
\label{eq_2bP}
\clos{\bar{P}}{2} = \frac{1}{2!} \, \sum \, \cbP{p_1} \cbP{p_2} \ket{v} \bra{v} \dbP{p_2} \dbP{p_1}
\end{equation}
reduces to the one of $N=2$ elementary bosons. A similar result is found for the closure relation for $N$ polaritons constructed on bosonized excitons or on Frenkel excitons.

\section{Discussion}
\label{sec_7_disc}

We end this paper by comparing and discussing the different expressions of closure relations for composite bosons we have obtained. Let us first list them for clarity.
\begin{itemize}
\item[(i)] The closure relation for $N$ elementary bosons is diagonal and has a $1/N!$ prefactor.
\item[(ii)] The closure relation for $N$ Frenkel excitons has the same form and the same $1/N!$ prefactor.
\item[(iii)] The closure relation for $N$ polaritons made of bosonized excitons has the same form and the same $1/N!$ prefactor.
\item[(iv)] The closure relation for $N$ polaritons made of Frenkel excitons has the same form and the same $1/N!$ prefactor.
\item[(v)] The closure relation for $N$ Wannier excitons has the same form but a $(1/N!)^2$ prefactor.
\item[(vi)] The closure relation for $N$ polaritons made of Wannier excitons has a more complicated form with non-diagonal terms in polariton operators, and prefactors which depend on the photon or exciton fractions in the polaritons of interest. This closure relation however reduces to the one of $N$ photons with its diagonal form and its $1/N!$ prefactors when all polaritons are seen as pure photons, while it reduces to the one of $N$ Wannier excitons, with its  $(1/N!)^2$ prefactor when all polaritons are seen as pure Wannier excitons.
\end{itemize}

The closure relation for Wannier excitons, is, in some sense, the most interesting one since it has the same simple diagonal form of elementary bosons, but a huge prefactor change, $(1/N!)^2$ instead of $(1/N!)$. This seems to indicate that the internal structure of composite bosons must be quite important for the closure relation prefactor. However, we should not be too fast in driving such a conclusion. Indeed, the fact that the closure relation for $N$ Frenkel excitons in contrast has the $1/N!$ prefactor of elementary bosons, just proves that for other composite bosons which are also made of two fermions, the closure relation prefactor may still be identical to the one of elementary bosons. For Frenkel excitons, we have shown that this is linked with the degree of freedom reduction occurring when constructing the relevant phase space for fermion pairs out of which these excitons are made. 

Further along this line, the fact that the closure relations for $N$ Frenkel excitons, for $N$ polaritons made of bosonized excitons and for $N$ polaritons made of Frenkel excitons all have the $1/N!$ prefactor of elementary bosons, proves that the composite nature of the bosons made of fermion pairs possibly mixed with photons, is not a determining factor for difference in closure relation prefactors. 

With respect to compositeness, polariton made of Wannier excitons which is linear combination of photons and free electron-hole pairs, corresponds to a far more complex composite object, because it can vary from pure photon to pure Wannier exciton; so that, depending on the relative weight of Wannier exciton in the polariton, we do have a composite quantum particle made of linear combination of photons and free electron-hole pairs or just a photon-like elementary particle. This leads to a closure relation having a far more complicated form, although fully compact in terms of polariton-exciton (or polariton-photon) overlaps.

To conclude, we have constructed the closure relation for $N$ polariton states, starting from the ones for subspaces corresponding to $p$ photons and $(N-p)$ excitons, with $0\leq p\leq N$, out of which these $N$ polaritons are made. When polaritons are constructed on Wannier excitons made of free electrons and holes, with a closure relation having a prefactor $(1/N!)^2$ instead of $1/N!$ as for elementary bosons, we find that the closure relation for $N$ polaritons depends explicitely on the weight of the exciton in the polariton for both its form and its prefactor. This is rather reasonable since, if all polaritons were pure photons, the closure relation prefactor should be $1/N!$, while if they all were pure excitons, we must recover the one for $N$ Wannier excitons with its $(1/N!)^2$ prefactor. In contrast, the closure relation for polaritons made of bosonized excitons or Frenkel excitons, simply is the one of $N$ elementary bosons.

The fact that, unlike all the other closure relations we have up to now obtained, the closure relation for $N$ polaritons made of Wannier excitons is non-diagonal in polariton operators, outlines the inherent complexity of the internal structure of Wannier exciton polaritons as composite bosons. Most notably, as for the lifetime of $N$-Wannier exciton states due to Coulomb scatterings~\cite{CBM_PRL04}, the complex structure of the closure relation for Wannier exciton polaritons is going to induce noticeable differences with elementary bosons in all physical quantities involving sum rules.


\begin{thebibliography}{13}
\expandafter\ifx\csname natexlab\endcsname\relax\def\natexlab#1{#1}\fi
\expandafter\ifx\csname bibnamefont\endcsname\relax
  \def\bibnamefont#1{#1}\fi
\expandafter\ifx\csname bibfnamefont\endcsname\relax
  \def\bibfnamefont#1{#1}\fi
\expandafter\ifx\csname citenamefont\endcsname\relax
  \def\citenamefont#1{#1}\fi
\expandafter\ifx\csname url\endcsname\relax
  \def\url#1{\texttt{#1}}\fi
\expandafter\ifx\csname urlprefix\endcsname\relax\def\urlprefix{URL }\fi
\providecommand{\bibinfo}[2]{#2}
\providecommand{\eprint}[2][]{\url{#2}}

\bibitem[{\citenamefont{Combescot and Betbeder-Matibet}(2005)}]{CBM_PRB05}
\bibinfo{author}{\bibfnamefont{M.}~\bibnamefont{Combescot}} \bibnamefont{and}
  \bibinfo{author}{\bibfnamefont{O.}~\bibnamefont{Betbeder-Matibet}},
  \bibinfo{journal}{Phys.\ Rev. B} \textbf{\bibinfo{volume}{72}},
  \bibinfo{pages}{193105} (\bibinfo{year}{2005}).

\bibitem[{\citenamefont{Combescot et~al.}(2008)\citenamefont{Combescot,
  Betbeder-Matibet, and Dubin}}]{CBMDub_PhysRep07}
\bibinfo{author}{\bibfnamefont{M.}~\bibnamefont{Combescot}},
  \bibinfo{author}{\bibfnamefont{O.}~\bibnamefont{Betbeder-Matibet}},
  \bibnamefont{and} \bibinfo{author}{\bibfnamefont{F.}~\bibnamefont{Dubin}},
  \bibinfo{journal}{Physics Reports} \textbf{\bibinfo{volume}{463}},
  \bibinfo{pages}{215} (\bibinfo{year}{2008}).

\bibitem[{\citenamefont{Fetter and Walecka}(1971)}]{Fetter_Walecka_book_71}
\bibinfo{author}{\bibfnamefont{A.}~\bibnamefont{Fetter}} \bibnamefont{and}
  \bibinfo{author}{\bibfnamefont{J.}~\bibnamefont{Walecka}},
  \bibinfo{journal}{Quantum theory of many-particle systems, (International
  series in pure and applied physics), McGraw-Hill, San Francisco}
  (\bibinfo{year}{1971}).

\bibitem[{\citenamefont{Martin and Rothen}(2004)}]{Martin_Rothen_book_04}
\bibinfo{author}{\bibfnamefont{P.}~\bibnamefont{Martin}} \bibnamefont{and}
  \bibinfo{author}{\bibfnamefont{F.}~\bibnamefont{Rothen}},
  \bibinfo{journal}{Many-body problems and quantum field theory: an
  introduction, (Texts and monographs in physics), 2nd ed., Springer, Berlin}
  (\bibinfo{year}{2004}).

\bibitem[{\citenamefont{Combescot
  et~al.}(2007{\natexlab{a}})\citenamefont{Combescot, Betbeder-Matibet, and
  Combescot}}]{CBMC_PRB07}
\bibinfo{author}{\bibfnamefont{M.}~\bibnamefont{Combescot}},
  \bibinfo{author}{\bibfnamefont{O.}~\bibnamefont{Betbeder-Matibet}},
  \bibnamefont{and}
  \bibinfo{author}{\bibfnamefont{R.}~\bibnamefont{Combescot}},
  \bibinfo{journal}{Phys.\ Rev. B} \textbf{\bibinfo{volume}{75}},
  \bibinfo{pages}{174305} (\bibinfo{year}{2007}{\natexlab{a}}).

\bibitem[{\citenamefont{Combescot and Betbeder-Matibet}(2004)}]{CBM_PRL04}
\bibinfo{author}{\bibfnamefont{M.}~\bibnamefont{Combescot}} \bibnamefont{and}
  \bibinfo{author}{\bibfnamefont{O.}~\bibnamefont{Betbeder-Matibet}},
  \bibinfo{journal}{Phys.\ Rev.\ Lett.} \textbf{\bibinfo{volume}{93}},
  \bibinfo{pages}{016403} (\bibinfo{year}{2004}).

\bibitem[{\citenamefont{Combescot and
  Pogosov}(2008{\natexlab{a}})}]{CP_arXiv_2_08}
\bibinfo{author}{\bibfnamefont{M.}~\bibnamefont{Combescot}} \bibnamefont{and}
  \bibinfo{author}{\bibfnamefont{W.~V.} \bibnamefont{Pogosov}},
  \bibinfo{journal}{arXiv} p. \bibinfo{pages}{0802.0145}
  (\bibinfo{year}{2008}{\natexlab{a}}).

\bibitem[{\citenamefont{Hopfield}(1956)}]{Hopfield_58}
\bibinfo{author}{\bibfnamefont{J.}~\bibnamefont{Hopfield}},
  \bibinfo{journal}{Phys. Rev.} \textbf{\bibinfo{volume}{122}},
  \bibinfo{pages}{1555} (\bibinfo{year}{1956}).

\bibitem[{\citenamefont{Combescot and Pogosov}(2008{\natexlab{b}})}]{CP_PRB08}
\bibinfo{author}{\bibfnamefont{M.}~\bibnamefont{Combescot}} \bibnamefont{and}
  \bibinfo{author}{\bibfnamefont{W.~V.} \bibnamefont{Pogosov}},
  \bibinfo{journal}{Phys.\ Rev. B} \textbf{\bibinfo{volume}{77}},
  \bibinfo{pages}{085206} (\bibinfo{year}{2008}{\natexlab{b}}).

\bibitem[{\citenamefont{Combescot
  et~al.}(2007{\natexlab{b}})\citenamefont{Combescot, Dupertuis, and
  Betbeder-Matibet}}]{CDupBM_EPL07}
\bibinfo{author}{\bibfnamefont{M.}~\bibnamefont{Combescot}},
  \bibinfo{author}{\bibfnamefont{M.~A.} \bibnamefont{Dupertuis}},
  \bibnamefont{and}
  \bibinfo{author}{\bibfnamefont{O.}~\bibnamefont{Betbeder-Matibet}},
  \bibinfo{journal}{Europhys. Lett.} \textbf{\bibinfo{volume}{79}},
  \bibinfo{pages}{17001} (\bibinfo{year}{2007}{\natexlab{b}}).

\bibitem[{\citenamefont{Abramowitz and Stegun}(1972)}]{AbramSteg}
\bibinfo{author}{\bibfnamefont{M.}~\bibnamefont{Abramowitz}} \bibnamefont{and}
  \bibinfo{author}{\bibfnamefont{I.~A.} \bibnamefont{Stegun}},
  \bibinfo{journal}{Handbook of Mathematical functions (Dover publications:
  New-York)}  (\bibinfo{year}{1972}).

\bibitem[{\citenamefont{Gradshteyn and Ryzhik}(1965)}]{GradshRyz}
\bibinfo{author}{\bibfnamefont{I.}~\bibnamefont{Gradshteyn}} \bibnamefont{and}
  \bibinfo{author}{\bibfnamefont{I.}~\bibnamefont{Ryzhik}},
  \bibinfo{journal}{Table of Integrals, Series, and Products, Fourth edition
  (Academic Press, New York)}  (\bibinfo{year}{1965}).

\bibitem[{\citenamefont{Landau and Lifchitz}(1966)}]{LandauLifMQ}
\bibinfo{author}{\bibfnamefont{L.}~\bibnamefont{Landau}} \bibnamefont{and}
  \bibinfo{author}{\bibfnamefont{E.}~\bibnamefont{Lifchitz}},
  \bibinfo{journal}{Quantum Mechanics (MIR editions: Moscow)}
  (\bibinfo{year}{1966}).

\end{thebibliography}

\end{document}